\documentclass[11pt,a4paper]{article}

\usepackage{fullpage}

\usepackage{amsmath}
\usepackage{amssymb}
\usepackage{amsfonts}
\usepackage{graphicx}
\usepackage{xspace}
\usepackage[hidelinks]{hyperref}
\usepackage{url}
\usepackage{booktabs}

\usepackage[USenglish]{babel}
\usepackage[utf8]{inputenc}

\newcommand{\fig}{Figure\xspace}

\newcommand{\eg}{e.g.\xspace}
\newcommand{\ie}{i.e.\xspace}
\newcommand{\etal}{et~al.\@\xspace}

\newcommand{\she}{she (or he)\xspace}

\title{On the State and Importance of Reproducible Experimental
  Research in Parallel Computing}

\author{Sascha Hunold, Jesper Larsson Tr\"aff\\
Vienna University of Technology\\
Faculty of Informatics, Institute of Informations Systems\\
Research Group Parallel Computing\\
Favoritenstrasse 16/184-5, 1040 Vienna, Austria\\
Email: \texttt{\{hunold,traff\}@par.tuwien.ac.at}
}

\date{}

\begin{document}

\maketitle

\begin{abstract}
Computer science is also an experimental science. This is particularly
the case for \emph{parallel computing}, which is in a
total state of flux, and where experiments are necessary to
substantiate, complement, and challenge theoretical modeling and
analysis. Here, experimental work is as important as are advances in
theory, that are indeed often driven by the experimental findings.  In
parallel computing, scientific contributions presented in research
articles are therefore often based on experimental data, with a
substantial part devoted to presenting and discussing the experimental
findings.  As in all of experimental science, experiments must be
presented in a way that makes reproduction by other researchers
possible, \emph{in principle}. Despite appearance to the contrary, we contend
that reproducibility plays a small role, and is typically not
achieved. As can be found, articles often do not have a sufficiently
detailed description of their experiments, and do not make available
the software used to obtain the claimed results. As a consequence,
parallel computational results are most often impossible to reproduce,
often questionable, and therefore of little or no scientific value. We
believe that the description of how to reproduce findings should play
an important part in every serious, experiment-based parallel
computing research article.

We aim to initiate a discussion of the reproducibility issue in
parallel computing, and elaborate on the importance of reproducible
research for (1)~better and sounder technical/scientific papers, (2)~a
sounder and more efficient review process and (3)~more effective collective
work.  This paper expresses our current view on the subject and should
be read as a position statement for discussion and future work. We do
not consider the related (but no less important) issue of the quality
of the experimental design.
\end{abstract}

\section{Introduction}
\label{sec:intro}

Even more so than for sequential computing systems, current parallel
architectures, their programming models and languages, and
corresponding software stacks are so diverse and so complex to model
that theoretical analysis of algorithms and applications needs to be
complemented by experimental evaluation and feedback.  Research in
parallel computing therefore relies heavily on computational
experiments as a driving force towards understanding new systems,
algorithms and applications. In some respects, the role of experiments
in parallel computing falls in the tradition of the evolving fields of
``experimental algorithmics'' and ``algorithm
engineering''~\cite{McGeoch12,Sanders09}, but have a much wider scope
by spanning a deep and difficult to control software stack, complex
applications, and underlying hardware in a constant state of flux.
With experimental results being so crucial it is of importance to ask
and to clarify whether the reported experiments really support the
wide-ranging conclusions drawn from the presented results (and if not,
what other purpose the extensive experimentation then serves). This
question has many interrelated facets, among which are:
\begin{itemize}
\item 
experimental strategy and methodology: is the experiment sound and
well designed?
\item
presentation and persuasion: are the results properly and
extensively reported, including statistics, positive and negative
results, sufficient parametric variation and coverage?
\item
correctness: are the results, tools and programs correct?
\item
trust: are the experiments trustworthy? 
Can they be backed up by more extensive data if so desired?
\item
reproducibility: are the results and outcomes reproducible, 
at least \emph{in principle}, by other, independent researchers?
\end{itemize}

The issue of trustworthiness is the one we will mostly consider here,
and argue that by paying more attention to the issue of
reproducibility, the trustworthiness of experimental work in parallel
computing can (and must) be improved such that the outcome of
experimental papers can serve their intended purpose of reference
and/or stepping stones for further research. This issue is currently
receiving much attention in other (natural) sciences, that in turn
rely heavily on computational experiments, but these discussions are
conspicuously absent in parallel computing. This is of course ironic
and unfortunate, insofar as so many other sciences rely on solid,
well-understood and trustworthy parallel and high-performance
computing environments: systems, run-time environments, numerical and
non-numerical libraries, simulation systems, etc.

With this paper we aim to start a discussion of the need for
reproducible research. We intend to back up this discussion with a
more thorough analysis of the state of affairs in our field in order
to corroborate the sometimes blunt claims that will be made in the
following. The paper is structured as follows. Section~\ref{sec:trust}
expands on the trustworthiness issue (in parallel computing).
Section~\ref{sec:reprod_research} summarizes articles that address the
necessity for reproducible experiments \emph{in other sciences} that
rely heavily on computational experiments.  Section~\ref{sec:practice}
briefly overviews some attempts and solutions for reproducibility and
provenance.  Section~\ref{sec:parallel-experiments} describe some of
the particularities of experiment-based research in parallel
computing.  We then highlight the challenges for reproducible research
by looking specifically to the field of parallel computing in
Section~\ref{sec:challenges}.  In Section~\ref{sec:testcase} we
discuss own experience in a case study to obtain reproducibility in
research on distributed scheduling algorithms.
Section~\ref{sec:conclusions} summarizes and outlines short term
approaches.

\section{Trustworthiness in Computational Research Papers}
\label{sec:trust}

In typical parallel computing research papers, scientists mostly
report on performance improvements (whether in time or other
objective) that originate from differential observations made in their
local environments.  In computational science, and more broadly in
computationally supported science, the focus is on the achieved
results, and less on the computational resources invested to achieve
those results.
In both cases, however, the reader has to trust the authors that
\begin{enumerate}
\item the (parallel) program (more broadly: the computational tool or
  system) under consideration was bug-free,
\item the experimental design was carefully chosen,
\item the data were properly gathered and processed, and that 
\item the final figures were generated using correct statistical analysis.
\end{enumerate}
These four requirements of trust are not only important to later
readers (the assumption is that any parallel computing paper is
intended to be read!); but also to the reviewers, who are the first to
attempt to objectively evaluate the experimental results and the
conclusions drawn from these.  An obvious (we believe) main 
problem here is that many articles that contain computational results pay
little attention to a sufficiently detailed description of their experiments.
Indeed, the reader is often left with an impression that the
presentation of experimental results is just an undesired necessity to
``prove'' that a proposed method works such that a success
story can be told. However, if researchers really want to explore
the subject further and possibly reimplement certain aspects of the given
paper, they discover a lack of experimental details.
As a result, one cannot extend on this research subject,
compare to alternative approaches, or simply verify the findings
independently, which is of crucial importance for any claimed,
new results of consequence.

A computational experiment can be seen as a chain of interdependent
actions to be performed. These actions form an ``experimental workflow'',
which usually consists of three components:
\begin{enumerate}
\item a problem instance creator (e.g., a workload generator),
\item an implementation of the algorithm that solves the given problem 
(the actual program/ experimental environment to study a phenomenon), and
\item a data analysis component 
(\eg, in the simplest case a script that plots experimental data).
\end{enumerate}

In our experience, many articles in parallel computing only provide fractions of
component~(1), a vague description of~(2) and a few graphs that
originated from component~(3), but which itself is unavailable.  Thus,
if scientists want to extend on published works, they will need to
fill in many blanks of the article.  Even if they contact the
original authors, they often experience that authors do not
remember the implementation details correctly (for all three
components), or even worse, that the implementations and experimental
data are no longer available.  Researchers will therefore have to do
the time-consuming job of reimplementing the experiment as it might
have been originally intended.  This not only slows down the individual 
research progress but also the scientific progress in this domain in general.

\section{The Need for Reproducible Research}
\label{sec:reprod_research}

A surprisingly large amount of research work on the subject of
reproducible research exists, many of which point to the absence
of reproducibility in computational sciences, \eg, life sciences.

Casadevall and
Fang, although from the field of Microbiology, characterize the reproducibility problem as
follows: ``There may be no more important issue for authors and
reviewers than the question of reproducibility, a bedrock principle in
the conduct and validation of experimental
science''~\cite{Casadevall:2010hi}.  Both authors point out that for
articles being acceptable in natural sciences ``the Materials and
Methods section should include sufficient technical information to
allow the experiments to be repeated''~\cite{Casadevall:2010hi}.

Roger D. Peng examined the state of reproducibility in computational
science and defined the spectrum of reproducibility which spans from
``full replication of a study'' to ``no
replication''~\cite{Peng:2011et}.  He concluded that ``addressing this
problem will require either changing the behavior of the software
systems themselves or getting researchers to use other software
systems that are more amenable to reproducibility. Neither is likely
to happen quickly; old habits die hard [...]''.
He also believes that ``the biggest barrier to reproducible research
is the lack of a deeply ingrained culture that simply requires
reproducibility for all scientific claims''~\cite{Peng:2011et}.
 
Victoria Stodden summarizes the problem as follows: ``It is impossible
to believe most of the computational results presented at conferences
and in published papers today. Even mature branches of science,
despite all their efforts, suffer severely from the problem of errors
in final published conclusions. Traditional scientific publication is
incapable of finding and rooting out errors in scientific computation,
and standards of verifiability must be
developed''~\cite{Stodden:2011uq}.  She concludes that ``making both
the data and code underlying scientific findings conveniently
available in such a way that permits reproducibility is of urgent
priority for the \emph{credibility of the
  research}''~\cite{Stodden:2011uq}.

A recent editorial in \emph{Nature Methods} addresses the problem by imposing
new, stricter reporting standards, especially stating that
``\emph{Nature Methods} will be requesting more information about the
custom software used to implement the methods we
publish''~\cite{Nature13}, challenging current habits in (parallel) computing.

To overcome the credibility problem, Jill P. Mesirov proposed a system
for ensuring reproducibility, which consists of two components: (1)
``a Reproducible Research Environment (RRE) for doing the
computational work'' and (2) ``a Reproducible Research Publisher
(RRP), which is a document-preparation system, such as standard
word-processing software, that provides an easy link to the
RRE''~\cite{Mesirov:2010hb}.  She concludes that ``we need simple,
intuitive ways to both capture and embed our computational work
directly into our papers. The value of such tools goes beyond mere
documentation. They will encourage the next generation of scientists
to become `active' consumers of scientific publications---not just
looking at the figures and tables, but running computational
experiments to probe the results as they read the
paper''~\cite{Mesirov:2010hb}.

In 2009, Fomel and Claerbout edited a special issue of the IEEE
Computing in Science and Engineering journal, discussing the problems of
lack of reproducibility in computation-based sciences, and proposing
methods and tools to overcome some of the
problems~\cite{Donoho09,FomelClaerbout09,PengEckel09}.  This special
has contributions by Roger Peng, Randall LeVeque, Victoria Stodden,
and David Donoho, and the editorial ends by asking: ``Before you
publish your next paper, please ask yourself a question: Have I done
enough to allow the readers of my paper to verify and reproduce my
computational experiments? Your solution to reproducibility might
differ from the those described in this issue, but only with a joint
effort can we change the standards by which computational results are
rendered scientific.''

A well-argued, dissenting opinion on the need for and feasibility of
reproducible research was recently offered by
Drummond~\cite{Drummond12}. 
He addresses the arguments for reproducibility raised by the Yale Law School Roundtable~\cite{yale-cise}.
Drummond believes that
people have ``different views of what replicability
means''~\cite{Drummond12}.
He distinguishes between three concepts: 
\begin{enumerate}
\item reproducibility: experiment duplication as far as possible,
\item statistical replicability: replicating the experimental results,
  which could have been produced by chance due to limited sample size,
\item scientific replicability: focusing on replicating the result
  rather than the experiment.
\end{enumerate}
He states that ``only Scientific Replicability has any real claim to
be a gold standard''~\cite{Drummond12}.  Drummond argues that
submission of data and code, which is targeted by the reproducibility
movement, would be counterproductive.  He claims that ``many papers
are uncited and others have only a few citations'' and so ``the
majority of code would not be used''~\cite{Drummond12}.  Drummond also
states that reviewers already have a high workload and it will be a burden
to them to look into data and code.  He also believes that we should
not enforce a single scientific method as a single such method does
not exist (cf.\ Feyerabend~\cite{feyerabend2010against}).  In addition, the problem of scientific
misconduct is discussed, which is often taken as motivation for requesting
reproducible research.  Drummond notes that scientific misconduct has
always been part of science and concludes that ``[i]f we were somewhat
more skeptical about the results of scientific experiments, cases of
misconduct would probably have much less of an
impact''~\cite{Drummond12}.

\section{Reproducible Experiments in Practice}
\label{sec:practice}

After motivating why reproducible experiments are needed in
computational sciences, we summarize some recent efforts for obtaining
reproducibility.

In 2005, Simmhan \etal published a survey on the state of data
provenance in e-Science.  Their primary goal was to ``create a
\emph{taxonomy of data provenance}
characteristics''~\cite{Simmhan:2005wc}.  This study mainly focused on
data provenance systems that use workflows to model scientific
processes.  In their taxonomy of provenance, replication
(reproduction) of an experiment is only one of several possible
applications of provenance.  Other uses are for example data quality,
attribution (copyright/ownership) or information (context for
interpretation of data).  This survey compares available workflow
systems made to support computations in natural and life sciences,
\eg, physics, astronomy, chemical sciences, earth sciences, or
biology.  We note here that computer science was not mentioned as a
possible application domain, although one might say that capturing
data from mostly deterministic sources such as computers should be easier than
collecting data about mice and chimpanzees.

In 2011, Gavish and Donoho published an article that introduced
\emph{verifiable computational research}. This form of research is
based on the combination of three concepts: ``verifiable computational
result (VCR), VCR repository and Verifiable Result Identifier
(VRI)''~\cite{Gavish:2011hv}.  The VCR represents ``a computational
result (\eg, table, figure, chart, dataset), together with the
metadata describing in detail the computations that created
it''~\cite{Gavish:2011hv}.  The VCR repository archives computational
results and should be accessible as Web-service.  The VRI can be a URL
(web address) that ``universally and permanently identifies a
repository''~\cite{Gavish:2011hv}.  The problem of long-term
re-executability of experiments is also discussed and in this context
the authors state that ``re-execution requires that the computing
environment originally used still be available and licensed on the
repository'', which is ''a tremendous difficulty for perpetuating
publications''~\cite{Gavish:2011hv}.  For these reasons, they deduce
that ``readable source code, its dependencies, and the actual run-time
values of input and output parameters [...] functions are what we
really need in order to understand how computational results were
generated''~\cite{Gavish:2011hv}.

Goble~\etal suggested that scientists should exchange so-called
``research objects'' rather than traditional articles.  A ``research
object bundles a workflow, together with provenance traces [...],
operational semantics, [...] version history, author attributions,
citation credit, license etc.''~\cite{Goble:2012wq}.  The article
summarizes possible benefits and drawbacks for scientists that choose
to share or not to share data and source code. They note that ``open
science and open data are still movements in their infancy'' and
``that the real obstacles are social''~\cite{Goble:2012wq}.  The
article concludes that ``the whole scientific community---from the
lab to the publisher and policy makers---needs to rethink and
re-implement its value systems for scholarship, data, methods and
software''~\cite{Goble:2012wq}.

DeRoure~\etal introduced the myExperiment
website\footnote{\url{http://www.myexperiment.org/}} that scientists
can use to share methods and processes~\cite{DeRoure:2010uv}.  Their
approach uses scientific workflows and focusses on ``multiple
disciplines (biology, chemistry, social science, music,
astronomy)''~\cite{DeRoure:2010uv}.  It is striking that
computer science in general (and parallel computing in particular) is
again missing in
this list.

As already mentioned, several workflow engines exist, but seem to be
primarily developed to support scientific processes in life and
biosciences, \eg, Taverna~\cite{Hull:2006hd} or
Kepler~\cite{Jaeger:2004tl}.

VisTrails is a provenance software for ``data exploration and
visualization through workflows''~\cite{Silva:vy}.  It is able to
``track changes made to workflows by maintaining a detailed record of
all the steps followed in the exploration''~\cite{Silva:vy}.
The ALPS project\footnote{Algorithms and Libraries for Physics
  Simulations} is an early adopter of VisTrails.  ALPS is a
collaborative project, in which researchers from different
laboratories contribute time and code.  The website states that the
main goal of ALPS is to ``increase software reuse in the physics
community.''\footnote{\url{http://alps.comp-phys.org}} The article by
Bauer~\etal introduces the different components of ALPS, one of them
is the ALPS VisTrails package~\cite{Bauer:2011cp}.  We point out that
every figure in the article by Bauer~\etal has a link to the
corresponding VisTrails workflow.
Thus, these workflows can be imported into VisTrails and used to
recreate the figures shown in the article.  The reader has now the
opportunity to modify the way the data is processed or to change the
view on the data, \eg, select a different scale of axes.  With such
approach the reader can actively verify and evaluate data produced by
others.
 
Another attempt to provide provenance and reproducibility is the tool
Sumatra~\cite{Davison:hh}.  It is layered on top of version control
systems such as Git or Mercurial and keeps track of computational
experiments\footnote{\url{http://packages.python.org/Sumatra/introduction.html}}.
In particular, it records (1)~the code, (2)~the parameter files, and
(3)~the platform used in the experiment. 
Scientists can tag an experiment with information
\emph{why} it was conducted and \emph{what} the outcome was.  Sumatra
is platform-independent (written in Python) and supports the
concurrent execution of different jobs using MPI (the Message Passing
Interface)~\cite{MPI-3.0}.

The Org-mode within the Emacs editor helps to accomplish reproducible
research~\cite{SchulteD11,Schulte:2012tv}.  Besides the note taking
capabilities, Org-mode supports literate programming.  The user can
embed source code (\eg, in C, Python, R, ..) into the \textit{org}
document. When an \textit{org} document is exported, each source code
block is evaluated. The result of such code block evaluation can be a
table or figure, which is inserted into the final document (\eg, a
\LaTeX\ or HTML document). Org-mode allows to define variables and to
pass values between code blocks.
The literate programming features of Org-mode are similar to the
functionality of Sweave.  Sweave allows researchers to embed R code
into \LaTeX\xspace documents~\cite{Leisch:2002}.

The publishing company Elsevier launched the Executable Paper Grand
Challenge\footnote{\url{http://www.elsevier.com/physical-sciences/computer-science/executable-papers}}
in 2011.  The winning Collage system has been used exemplary for
special issues of Elsevier journals.  Elsevier's Collage system is
comparable to Org-mode as it allows to embed code blocks and data
items into an executable paper.

In his dissertation Guo addresses the problem of reducing the
complexity of research programming that often stems from the
programmers' burden of dealing with data management and provenance
issues~\cite{Guo:2012tc}.  He presents several software tools to
facilitate the day-by-day routines of a scientific programmer such as
data cleaning, data analysis, or experiment deployment and management.

\section{The State of Experiments in Parallel Computing}
\label{sec:parallel-experiments}

Towards providing tools in the form of algorithms, languages,
libraries and systems for scientific and other applications to use,
and towards understanding these algorithms, languages, libraries and
systems themselves, parallel computing very often relies on experimental
results. A typical, experimental parallel computing paper describes an
algorithm or system, makes claims that it improves in all or certain
aspects over a previous algorithm or system, not rarely one's own, and
purports to show this by a number of experiments. If carefully
implemented, the claims may indeed count as at least well supported,
and the reader will believe and be able to build on the results so
established.

On closer look, this ideal, template setup is, however, often
problematic. Due to ``space limitations'', the algorithm---especially
if part of a larger, complex application or system---may not be
described in enough detail to allow reimplementation, neither for
rechecking the claims nor for using the result. This would not be a
problem if the actual program would be available, or at least enough
supplementary material to facilitate reimplementation without
difficulty. We contend that this is rarely the case, and will later
(not this paper) support
the claim by examining a selection of recent (and past) parallel
processing conferences and journals.
In the cases where the actual implementation as
a program is not available for inspection, the next issue is whether
the program is indeed correct (bug-free) as claimed. Related to this
is the case where the algorithm/program is only applicable to special
cases of inputs (the notorious powers-of-two); such restrictions in
generality are rarely stated, but do limit claims of universality: an
implementation that actually does work (correctly) in all intended
cases may well be less impressive (in terms of speed, overhead,
scalability, energy consumption, or whatever metric is used in the
evaluation) than the limited implementation that has actually (perhaps
inadvertently) been used. But the case remains: without a precise
description and perhaps access to at least parts of the actual code,
it is not possible for the reader to judge.

Parallel systems are more complex than ``traditional'', sequential
computing systems in at least one extra dimension: many, sometimes
very many, processors or cores with complex interconnections via
shared-memory resources or communication networks. This is in
addition to the complexities of modern processors, where the
utilization of the memory system (cache-hierarchy) and instruction set
can make a gigantic impact on the performance (whether measured as
time or energy or something else). Thus, to judge and reproduce a
given result, these system parameters have to be stated very
carefully.  Also compiler versions and settings can make huge
differences. There is in the parallel computing field a positive
tendency towards reporting some of these factors and parameters, but
whether this information is sufficient for evaluating scientific
results should be examined by some case studies.

A particular problem in parallel and high-performance computing is the
ephemeral, unique and even exclusive character of the systems, \eg,
many (most) systems on the TOP500\footnote{\url{http://www.top500.org}} have
restricted access and have a limited lifespan. Therefore, the
possibility to test and measure a piece of software (an algorithm, a
library, an application) is for most researchers a once-in-a-lifetime
opportunity. This is a fact that should be taken into account when
judging the character of results on such systems.

Other problems with experiments in parallel computing, especially for
large-scale systems, are the effects of operating system ``noise'',
interference from network and file-systems, other users, etc., that
are very difficult to control and often ignored (sometimes out of
ignorance), but may heavily distort or bias results. Again, more
attention to reproducibility would automatically lead to more concern
with these issues. 

Tools like libraries (MPI) and compilers (OpenMP) are sometimes used
to gather timing and other results. It is assumed, but most often not
questioned that such performance results are accurate and
reproducible. Recent experience (with OpenMP) has shown that this is
not the case: the built-in timing facility, \texttt{omp\_get\_wtime()},
can be heavily dependent on the number of started threads,
in a counter-intuitive way; also timing results showed higher
variation than naively expected. Solid experimental work must be aware
of such features. An issue here is that the OpenMP standard (and
similar standards) does not prescribe any specific behavior that an
implementation must fulfill.

Experimental parallel computing research, when focussed on algorithms
and applications, is often concerned with showing relative
improvements and scalability. A standard measure is the ``speed-up''
achieved relative to some baseline~\cite{RauberRunger10}. Without
explicitly stating what this baseline is (and sometimes a discussion
of whether the baseline is meaningful would also seem in order), the
results can be grossly misleading and, indeed, not reproducible. Often
such statements are missing, inadvertently or by intent. These
problems are known and have been
discussed~\cite{Bailey92,Bailey09,HagerWellein11}. Bad experimental practice
also in not infrequent cases lead to censoring of results to
highlight only the good things.

In order to address these issue raised in this section, we will now
look at potential challenges that we face when attempting to improve
reproducibility in parallel computing.

\section{Challenges for Parallel Computing Research}
\label{sec:challenges}

Comparing algorithms or scientific approaches via experiments or
simulations is standard practice in parallel computing.  For example,
every article of Issue No.\ 1, 2013 in Volume 24 of the IEEE
Transactions on Parallel and Distributed Systems\footnote{TPDS was
  chosen since it is a well established and highly respected journal
  in the domain.  The issue was selected as it was the latest issue as
  of the time of working on the present article.} (TPDS) justifies the
scientific contribution through experiments or simulations.
Unfortunately, the current standard for publishing articles has
shortcomings if a reader wants to reproduce the results.  An obvious
problem is that the source code is usually not part of an article.  In
fact, only a small number of articles contain a link to the source
code. In the particular case of the mentioned issue of TPDS, we could
only find one paper containing a link to the corresponding source
code. Naturally, a single issue of a specific journal is
not representative for the entire domain of parallel
computing. However, we contend that it supports our statement that
reproducibility is an issue that should not be overlooked.

The question is then how to approach the problem. Answering this
question will not be easy as there are technical, social, and
political constraints and implications.  Nevertheless, in
this paper we would like to name goals to improve reproducibility in
parallel computing.

\subsection{A Clear Objective for Reproducibility}

First, we need to clarify which level of reproducibility we expect
or is possible in our context. Drummond points out that there is
no consensus on what reproducibility of an experiment means.

In the context of parallel computing we are mostly interested in what
Drummond called ``scientific replicability''.  So, we would like to
reproduce and verify the scientific outcome presented in articles
rather than reproducing the exact same numbers of a previous
experiment.  Nonetheless, numerical reproduciblity as considered by Demmel and
Nguyen~\cite{DemmelN13a} is equally important for large-scale
experiments and might become a prerequisite for obtaining scientific
replicability in our context.

As previously mentioned, ``scientific replicability'' is especially
important in parallel computing where certain experiments require
specific hardware to be conducted. For example, only few scientists
have the opportunity to execute a parallel algorithm on a large number
of processors on the latest supercomputer from the TOP500 list.  Yet,
scientists should have the possibility to conduct similar experiments
at smaller scale to verify findings, to analyze the source code if
needed, or to access experimental data used in articles.

\subsection{Verifying the Reproducibility Problem}

As we now know what we are expecting when we demand reproducible
results in parallel computing, we should evaluate the state of
reproducibility in this domain. We need an unbiased, scientifically
sound survey whether experimental results in parallel computing are
reproducible to our standards or not. Surely, from our
experience we know that reproducing experimental results is hard and
often impossible. However, a personal feeling cannot replace a clean,
objective, scientific study. In discussions with other researchers in
our field we noticed that the problem of experimental reproducibility
in parallel computing is not always a major concern.  We have often
heard from others (cf. also Drummond) that research that is not reproducible will eventually vanish
and its lasting significance will be low. Thus, according to this
opinion, enforcing stricter rules such as providing data and source
code will only put a burden on reviewers. This argument coincided with
Drummond's statements in~\cite{Drummond12}.  The problem with this
argumentation is that it takes time to invalidate findings, most often
several weeks of work of PhD students to figure out that something is
not quite right.

The basic problem of today's science world is the paper pressure on
researchers---``we need to publish something''. This is a global
phenomenon.  Productivity of researchers is hardly
measurable in a single index, yet, universities, various rankings,
politics, etc. are desperate to compare the scientific impact.  Thus,
conferences and journals experience a flood of research articles
claiming all sorts of advances.  We argue that the number of articles
submitted for revision could significantly be reduced if we increase
the standards for reproducibility. Nonetheless, we agree with Drummond
that no specific ``method of science'' should be enforced. So,
researchers must have the freedom to submit research articles
in whatever flavor. However, publishing computational results
without any form of verifiability should count as unacceptable.

\subsection{Stricter Publishing Rules}

A first step would be if editors would enforce authors to release all
experimental details as supplementary material, although the burden
will be high compared with the status quo. This could also be extended
beyond publishing a regular journal articles, \ie, also peer-reviewed
conferences could apply such best practices.  The ACM Journal on
Experimental Algorithmics is one journal that tries to encourage
authors to make their programs and testbeds available.  The editors of
the journal state that ``[c]ommunication among researchers in this
area must include more than a summary of results or a discussion of
methods; the actual programs and data used are of critical
importance''\footnote{\url{http://www.jea.acm.org/about.html}}.

These stricter publishing rules should be applied to all research
articles that claim scientific contributions based on experimental
data. A main issue to consider here is where to draw the line between
theory and (experimental) practice. Of course, there are also legal
restrictions for researchers to release certain data. Such issues need
to be taken into account when implementing new publishing guidelines.
In the end, we need to ask ourselves how valuable research papers are
that claim scientific improvements from experimental data
gathered using proprietary, undisclosed software and hardware.

We additionally contend that the review process in general needs to
improve, especially for experimental results. 
The reviewer has to take the presented results at face value. Rarely, an effort
is made to check the results, and as we have explained, this often
not possible. 
Very often, submitted papers in (experimental) parallel
computing compare experimental results based on a shaky statistical
analysis. We often see the mean of absolute run-times of some
algorithms, without having any information about the number of trials
or the dispersion of values. Reviewers in our domain barely verify
whether the results are statistically significant. In addition, since
the experimental data is not available as part of supplemental data,
research papers can hide important problems or increase certain
effects almost arbitrarily~\cite{Bailey09,hager_blog}.  Best practices
are known for years~\cite{jain1991,leboudec2010performance} and should
therefore be enforced.

\subsection{Improving Software Tools}

There are also some technical means that might help towards making
computational results reproducible, in particular
better software tools.  

Computer science, especially parallel computing, has different
requirements than most other computational sciences.  Here we are
often concerned with finding the best algorithm to solve a given
problem. We use metrics like GFLOP/s or speed-up that are based on the
actual run-time of a program. Thus, we are interested in measuring the
absolute run-time of an experiment. This is in contrast to other
computational sciences, \eg, computational biology or chemistry, where
researchers are mostly interested in the result of a computation
rather than the execution time.

We believe that computer science research misses tools that support
scientists in obtaining reproducible results and that introduce only little
noise into the experiment (as we measure execution time).  For a wider
adoption in computer science research, GUI\footnote{Graphical User
  Interface}-driven workflow engines seem to be less suited.  These
tools are often too heavyweight and may slow down the actual program
execution.

It could also be the case that most building blocks for a
reproducibility framework have already been developed, \eg, Git,
Python, Sumatra, Org-mode, make, cmake, autotools, etc.  However, we
lack expertise how to combine them properly to achieve
reproducibility.

What we are missing is a better understanding what we need to record,
at which experimental detail, and how to do that efficiently.
Peng introduced the ``Reproducibility Spectrum'' that ranges from not
reproducible (having only the publication) to full replication of
experiments~\cite{Peng:2011et}.  As said before, we are mostly
interested in scientific replication. Yet, how many experimental
details are needed to achieve this goal is an open question, which we
are going to address in future research.

\section{Reproducibility Test Case --- Distributed Scheduling}
\label{sec:testcase}

We would like to share our experiences that we have made while trying
to achieve reproducibility in our own work. The original research
question was how to maximize the steady-state task throughput (in
operations per second) on a computational grid.  We developed a fully
distributed algorithm that schedules Bag-of-Tasks applications onto a
computer grid~\cite{bertin2011}.  We developed a simulator on top of
SimGrid~\cite{CasanovaLQ08} to evaluate whether our algorithm
converges.
In order to share code and experimental data with other researchers
(and reviewers), we created a webpage providing the source code of the
simulator and the R code used to analyze the
data\footnote{\url{http://mescal.imag.fr/membres/arnaud.legrand/distla_2012/}};
this website also provides additional information such as
visualizations.  In order to evaluate our scheduling algorithm through
simulation in many different test cases, we ran a large number of
experiments (parameter-sweep).
Altogether, answering the research question whether our fully distributed
scheduling algorithm optimizes the total throughout required thousands
of such simulations.

A computational experiment often consists of a loosely coupled
sequence of commands that one needs to execute.  The term ``loosely''
is important here as it distinguishes them from reproducible
experiments, for which dependencies are explicitly stated. From
experience we know that these loosely coupled experiments occur
frequently in our domain.  For a single researcher, this style of
experimentation works well in the beginning of a project, but gets
troublesome later, especially if \she wants to collaborate with other
researchers.  Typical questions that will arise---and arose for
us---are for example: (1)~How did I/you call this script? (2)~Which
file do I/you need to edit to set up a path or parameter? (3)~Which
version of software 'X' have I/you used in the original analysis?
Thus, it happens frequently that one cannot reproduce her (or his) own
work.

\begin{figure}[t]
\centering
\includegraphics[width=\linewidth]{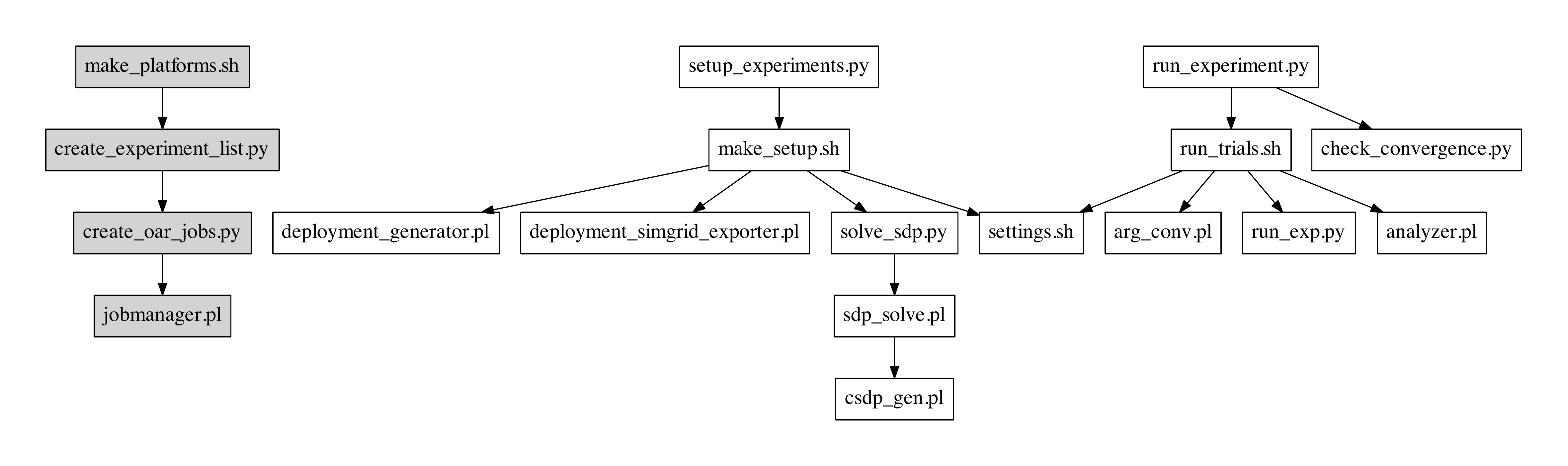}
\caption{\label{fig:depgraph}Steps to run a batch of simulations (left). 
Dependency graph for setting up and running a single experiment (right).}
\end{figure}

\begin{figure*}[t!]
\centering
\includegraphics[width=\linewidth]{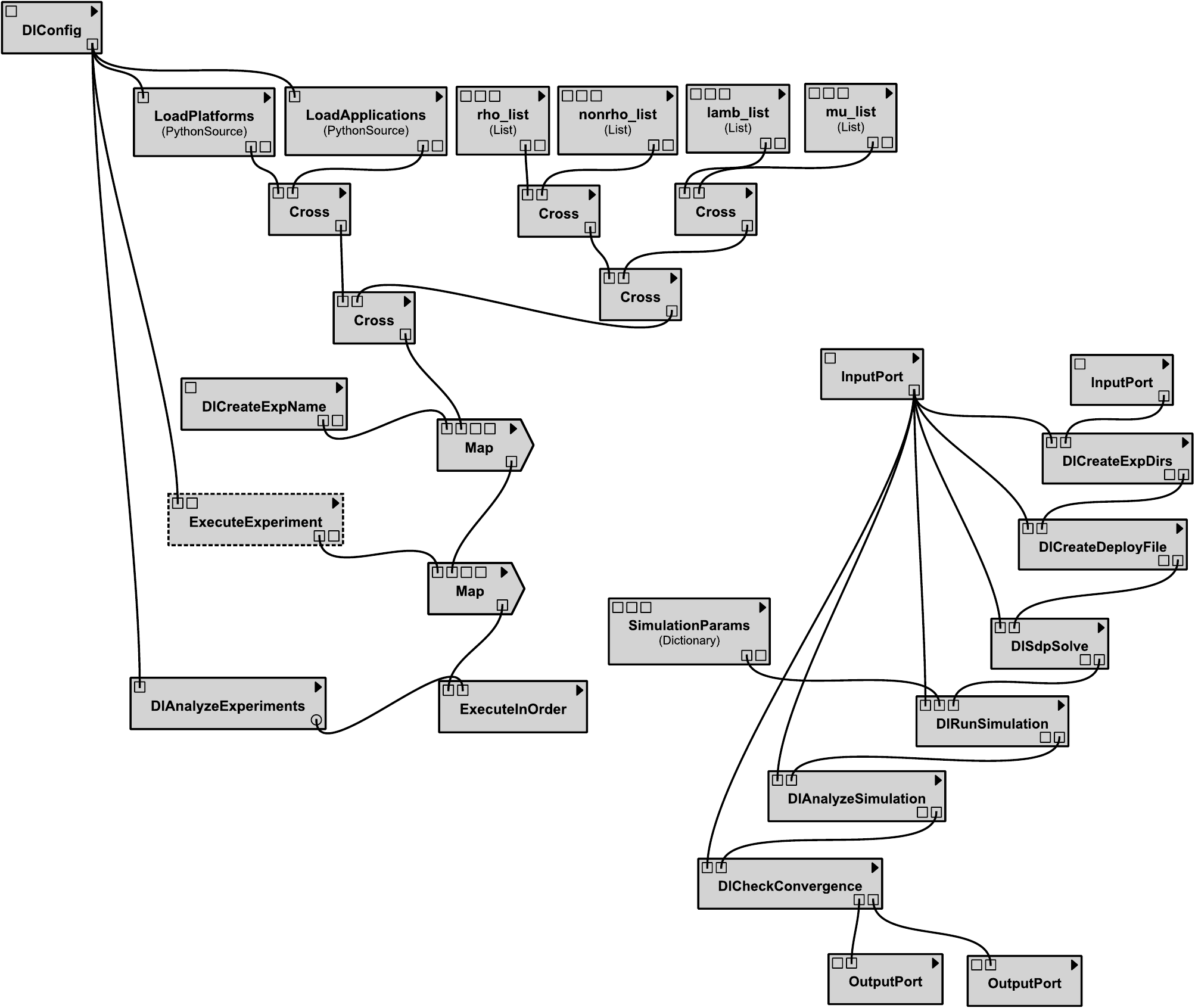}
\caption{\label{fig:vistrails}VisTrails workflow for running a batch
  of simulations (left) with subgraph of node ``ExecuteExperiment''
  (right).}
\end{figure*}

\fig~\ref{fig:depgraph} (left) shows the right sequence of our scripts
to execute a batch of simulations.  Each script takes a long list of
parameters that define a particular simulation run, \eg, the number of
computer nodes, the number of applications, the computational amount
of each task, etc. The right-hand side of \fig~\ref{fig:depgraph}
depicts the call graph to setup and to execute a single experiment.
Now, having structured the scripts nicely for the present publication,
the order of steps to conduct the simulation study seems obvious.
Before, we had a situation where all scripts were placed in one
directory. In such case, it was not obvious for another person to
figure out the correct order to call the scripts.  To help others (and
primarily ourselves) to rerun our (own) experiments, we wanted to
capture our experimental process in a workflow.  VisTrails seemed to
be a good starting point as most of our scripts were written in
Python, which is the language of VisTrails.  The resulting VisTrails
workflow is shown in~\fig~\ref{fig:vistrails}.

The main advantage of the new VisTrails workflow was that the time to
setup a simulation run on a colleague's machine could be significantly
reduced.  In addition, merging single scripts into one larger workflow
also removed redundancy as each workflow module\footnote{A module is a
  workflow node.} was now accessing shared parameters.  VisTrails made
it easy to generate parameter combinations as it provides modules to
manipulate parameters, \eg, applying a cross product or filtering
elements (see~\fig~\ref{fig:vistrails}).
Yet, the VisTrails approach also entailed disadvantages for us.  As we
were relatively new to VisTrails (but not new to Python), it took us
surprisingly long to obtain the final workflow of the experiment.
Connecting workflow nodes was error prone as all modules are written
in Python and linking incompatible variables could only be caught at
runtime due to the dynamically typed programming language.  Further,
the workflow approach makes debugging and writing single modules
harder.  A programmer usually builds a workflow by adding modules
sequentially from top to bottom. Thus, adding a buggy module will be
caught only when the module is executed, and this module often happens
to be the last one in the graph.  VisTrails tracks changes to the
workflow but does not monitor external code changes.  We wrapped most
of our existing script in VisTrails modules to make them available for
the workflow engine.  So, changes to our source code were not recorded
by VisTrails, and we had to use an external version control system
(Git).  At the time of working with VisTrails, it supported the
parallel execution of workflows on workflow level, \ie, one could
execute several workflows concurrently.  A more fine-grained parallel
execution (\eg, a parallel implementation of a ``Map'' node) would be
needed to speed up the execution of a single workflow.

The VisTrails approach entailed another disavantage\footnote{We note
  that not the particular software VisTrails entailed disadvantages
  for us, but rather the concept of having a centralized workflow
  engine.}.  In our original experimental procedure (without workflow
engine), we ran batches of simulation experiments on
Grid'5000\footnote{\url{https://www.grid5000.fr}} to reduce the time
to complete the study.  Developing VisTrails modules that connect to
Grid'5000 via SSH, submit jobs, and monitor the job execution would
have required additional programming effort.  Morever, such VisTrails
workflow would be too specific to our environment and hardly
executable on another machine that has no access to Grid'5000.  Hence,
we decided to limited the executability of our VisTrails workflow to
the user's machine.

The portability of the VisTrails workflow was another problem as
VisTrails only eases the execution of scripts.  It does not simplify
the necessary software configuration, \ie, most of our Perl or Python
scripts call other programs such as time, awk, R, or DSDP5.  In
addition, our simulator needs to be compiled with a specific version
of the SimGrid library.  To overcome these problems, we created a
Virtual Machine (VM) using
VirtualBox\footnote{\url{https://www.virtualbox.org/}} and installed a
GNU Debian Linux (Wheezy) with all required packages (Python, gcc,
Perl, DSDP5, gnuplot, SimGrid, etc.).  Now, our workflow became
executable by others; users could simply start the VM and either run
VisTrails or our original scripts.  To reduce the simulation time, we
installed a simple job manager to take advantage of all processor
available to the virtual machine by spawning as many simulation
processes as there are cores.  Clearly, the disadvantage is now that a
researcher has to download the virtual machine image (which can be
large depending on the case study).  Currently, it seems unpractical
to provide a VM machine as supplementary material of research
articles. In addition, a virtual machine image might not be startable
in five years from now.  Yet, we believe that this might not be our
biggest concern as the computing world is so rapidly changing that
experimental results could be obsolete in five years, in contrast to theoretical
findings.

In summary, we could improve the reproducibility of our experiments
thanks to VisTrails.  However, several issues have to be addressed in
future work.  First, we have no evidence whether our work is truly
reproducible by others or not; this could only be evaluated by a third
party. Second, certain technical issues have to be solved, \eg, is it
really necessary to create a virtual machine?  Third, in the presented
case study the execution time was not of primary interest for the
experiment.  As performance measurements are of major concern in our
research domain, we will need to investigate such
experiments. Nonetheless, we have learned that reaching for
reproducibility is a time-consuming process, which requires high
dedication and stamina.


\section{Conclusions}
\label{sec:conclusions}

We have aimed to initiate a discussion of the state and the importance
reproducibility in parallel computing. The aim is to examine standards
for good, trustworthy, and ultimately enlightening and useful
experimental work in this field of computer science. We contend that
more attention to doing and reporting research that is, in principle,
reproducible by other researcher will improve the quality of the
experimental research conducted. We have discussed some tools and
approaches that might help in this direction, and reported on our own
attempt to make our research reproducible. Much further work is
required, and we will extend this study and explore other tools. We
also intend to conduct a broader study and analysis of current
experimental practices in parallel computing.

\bibliographystyle{abbrv}
\bibliography{reprod}

\end{document}